\documentclass[aps]{revtex4}

\usepackage{graphicx}

\usepackage{amsmath}

\def\kslash{k\kern-.5em\slash}
\def\Pslash{P\kern-.5em\slash}
\def\pslash{p\kern-.5em\slash}
\def\qslash{q\kern-.5em\slash}
\def\dslash{\partial\kern-.5em\slash}

\begin{document}

\title{A $q$-deformed NJL model at finite temperature: \\chiral symmetry restoration and pion properties}
\author{V. S. Tim\'oteo}
\affiliation{Centro Superior de Educa\c{c}\~{a}o Tecnol\'{o}gica, Universidade Estadual de Campinas, 13484-370, Limeira, SP, Brazil}
\author{C. L. Lima}
\affiliation{Instituto de F\'{\i}sica, Universidade de S\~{a}o Paulo, CP 66318, 05315-970, S\~{a}o Paulo, SP, Brazil}

\begin{abstract}
We review the implementation of a $q$-deformed fermionic algebra in the
Nambu--Jona-Lasinio model (NJL). The gap equations obtained from a deformed
condensate as well as from the deformation of the NJL Hamiltonian are
discussed. The effect of both temperature and deformation in the chiral
symmetry restoration process as well as in the pion properties is studied.

\end{abstract}
\maketitle

\section{Introduction}

Chiral symmetry breaking is an important phenomena in hadron physics 
and is of fundamental importance for hadron properties. The difficulties
involved in obtaining low-energy properties directly from QCD, the fundamental theory
of strong interactions, have motivated the construction of effective models.
Due to its simplicity and effectiveness in describing hadrons at low energies, 
the NJL model \cite{NJL} has become the most familiar effective model for strong
interactions. 

The standard scenario for spontaneous chiral symmetry breaking is the arising
of a quark condensate. In the NJL model the condensates appear when the
strength of the contact interaction exceeds a critical value, separating the
Wigner-Weyl and Nambu-Goldstone realizations of chiral symmetry. The
condensate $<\bar{q}q>$ is the order parameter. 

Apart from giving a reasonable description of the light mesons, the NJL model
is also very useful for studying the chiral symmetry breaking process as well as its
restoration at finite temperature \cite{ASYA,VW}. The NJL model is in addition very 
suitable for testing new ideas. In previous works, we have investigated the effect of $q$-deformation in the chiral symmetry breaking process within the context of the NJL model. In particular, we  observed how the condensate is affected by the deformation. As a direct consequence, the dynamical quark mass, the chiral symmetry breaking and restoration processes are accordingly affected \cite{VST,CLL}.

The aim of this work is to study the effect of both temperature and deformation
in the chiral symmetry restoration, and evaluate the pion properties in a $q$-deformed NJL model at finite temperature.
This paper is organized as follows. Sec. II introduces the $q$-deformed fermionic algebra which will be used along
this work, and shows how the deformed algebra is implemented in the
NJL model. In Sec. III, we present the investigation of the chiral symmetry restoration with both
finite temperature and $q$-deformation. Finally, Sec. IV contains our main conclusions.

\section{$q$-Deformation in the NJL model}
The Nambu--Jona-Lasinio model was first introduced to describe the nucleon-nucleon interaction via a four-fermion contact interaction. Later, the model was extended to quark degrees of freedom becoming an effective model for quantum chromodynamics. Details of the NJL model and the Bogoliubov-Valatin variational approach can be found in review works  \cite{VW,K}.

In this section we discuss the implementation of the deformed algebra to the NJL model. 
The $q$-deformed fermionic algebra \cite{gal} that we shall use is based in
the work of Ubriaco \cite{ubri}, where the thermodynamic properties of a many
fermion system were studied. In the
construction of a $q$-covariant form of the BCS approximation \cite{trip}, it
was shown that the creation and annihilation operators of the $su_{q}\left(
2j+1\right)  $ fermionic algebra are given by
\begin{equation}
\mathcal{A}_{jm_j}=a_{jm_j}\prod_{i=m_j+1}\left(  1+Qa_{ji}^{\dagger}a_{ji}\right)  ,
\label{A}
\end{equation}
\begin{equation}
\mathcal{A}_{jm_j}^{\dagger}=a_{jm_j}^{\dagger}
\prod_{i=m_j+1}
\left(1+Qa_{ji}^{\dagger}a_{ji}\right)  ,
\label{Ad}
\end{equation}
where $Q=q^{-1}-1,$ $j=1/2$ and $m_j=\pm1/2$. As discussed in \cite{VST}, the first consequence of the above
deformation is that only the operators corresponding 
to $m_j=-\frac{1}{2}$ are
modified, meaning that only negative helicity quarks (anti-quarks) operators
will be deformed since in the NJL model we deal with quarks (anti-quarks)
creation and annihilation operators. 

We have two different approaches to
obtain a new gap equation. The first one consists in to perform the
$q$-deformation in the condensate, which is a part of the self-consistent
equation for the dynamical mass. In the second one the $q$-deformation is
performed in the NJL Hamiltonian, and we use the Bogoliubov-Valatin procedure
to obtain the new gap equation. We now turn to a detailed discussion of these
two different approaches.

\subsection{Deforming the condensates}

To obtain the deformed gap equation we work with the BCS-like vacuum in the standard Bogoliubov-Valatin variational approach
\begin{equation}
\left|  NJL\right\rangle =\prod_{\mathbf{p},s=\pm1}\left[  \cos\theta
(p)+s\sin\theta(p)b^{\dagger}(\mathbf{p},s)d^{\dagger}(-\mathbf{p},s)\right]
\left|  0\right\rangle ~, 
\label{bcs}
\end{equation}
which, for a given momentum $\mathbf{p}$, is expanded as
\begin{align}
\left|  NJL\right\rangle  &  =\cos^{2}\theta(p)\left|  0\right\rangle
\nonumber\\
&  +\sin\theta(p)\cos\theta(p)b^{\dagger}(\mathbf{p},+)d^{\dagger}
(-\mathbf{p},+)\left|  0\right\rangle \nonumber\\
&  -\sin\theta(p)\cos\theta(p)b^{\dagger}(\mathbf{p},-)d^{\dagger}
(-\mathbf{p},-)\left|  0\right\rangle \nonumber\\
&  -\sin^{2}\theta(p)b^{\dagger}(\mathbf{p},-)d^{\dagger}(-\mathbf{p}
,-)b^{\dagger}(\mathbf{p},+)d^{\dagger}(-\mathbf{p},+)\left|  0\right\rangle ~.
\label{bcsexp}
\end{align}

The quark fields are expressed in terms of $q$-deformed creation and annihilation operators as
\begin{equation}
\psi_{q}(x,0)=\sum_{s}\int\frac{d^{3}p}{\left(  2\pi\right)  ^{3}} \left[
\mathcal{B}(\mathbf{p},s)u(\mathbf{p},s)e^{i\mathbf{p\cdot x}}+ \mathcal{D}
^{\dagger}(\mathbf{p},s) v(\mathbf{p},s)e^{-i\mathbf{p\cdot x}}\right]  ,
\label{bd1}
\end{equation}
where the $q$-deformed quark and anti-quark creation and annihilation
operators $\mathcal{B}$, $\mathcal{B}^{\dagger}$, $\mathcal{D}$, and
$\mathcal{D}^{\dagger}$, are expressed in terms of the non-deformed ones
according to Eqs. (\ref{A}) and (\ref{Ad})
\begin{align}
\mathcal{B}_{-}  &  = b_{-}\left(  1+Qb_{+}^{\dagger}b_{+}\right)  \;\;\; ,
\;\;\; \mathcal{B}_{-}^{\dagger}=b_{-}^{\dagger}\left(  1+Qb_{+}^{\dagger
}b_{+}\right)  ,\\
\mathcal{D}_{-}  &  = d_{-}\left(  1+Qd_{+}^{\dagger}d_{+}\right)  \;\;\; ,
\;\;\; \mathcal{D}_{-}^{\dagger}=d_{-}^{\dagger}\left(  1+Qd_{+}^{\dagger
}d_{+}\right)  , \label{bm}
\end{align}
\begin{align}
\mathcal{B}_{+}  &  = b_{+} \;\;\; , \;\;\; \mathcal{B}_{+}^{\dagger}
=b_{+}^{\dagger},\\
\mathcal{D}_{+}  &  = d_{+} \;\;\; , \;\;\; \mathcal{D}_{+}^{\dagger}
=d_{+}^{\dagger}, \label{bp}
\end{align}
where $+$ $(-)$ stands for the positive (negative) helicity and the notation has been simplified: $(\mathbf{p},s)\rightarrow {s}$. 
We would like to note that, as discussed
in Ref. \cite{trip}, the deformed vacuum differs from the non-deformed one
only by a phase and, therefore, the effects of the deformation comes solely
from the modified field operators. Additionally, the $q$-deformed NJL
Lagrangian, constructed using $\psi_{q}$ instead of $\psi$, is invariant under
the quantum group $SU_{q}(2)$ transformations. This can be seen by using the
two-dimensional representation of the $SU_{q}(2)$ unitary transformation given
in Ref. \cite{ubri}.

The deformed gap equation is
\begin{equation}
m=-2G\left\langle \overline{\psi}\psi\right\rangle _{q}, \label{qmdy}
\end{equation}
where $\left\langle \overline{\psi}\psi\right\rangle _{q}$ is the $q$-deformed
condensate calculated using the BCS-like vacuum Eq. (\ref{bcs}),
\begin{align}
\left\langle \overline{\psi}\psi\right\rangle _{q}  &  =\left\langle
NJL\left|  \bar{\psi}_{q}\psi_{q}\right|  NJL\right\rangle \nonumber\\
&  =\left\langle \overline{\psi}\psi\right\rangle +\left\langle NJL\left|
\mathcal{Q}\right|  NJL\right\rangle ,
\end{align}
where $\left\langle \overline{\psi}\psi\right\rangle $ is the non-deformed
condensate and $\left\langle NJL\left|  \mathcal{Q}\right|  NJL\right\rangle $
represents all non-vanishing matrix elements arising from the $q$-deformation
of the quark fields. The contribution of these $q$-deformed matrix elements
is
\begin{equation}
\left\langle NJL\left|  \mathcal{Q}\right|  NJL\right\rangle =Q\int
\frac{d^{3}p}{\left(  2\pi\right)  ^{3}}\left[  \sin2\theta(p)-\sin
2\theta(p)\cos2\theta(p)\right]  .
\end{equation}

The calculation of the deformed condensate will be performed in a similar way
as in the usual case. It requires also a regularization procedure since the
NJL interaction is not perturbatively renormalizable. For reasons of
simplicity a non-covariant trimomentum cutoff is applied arising
\begin{equation}
\left\langle \overline{\psi}\psi\right\rangle _{q}=-\frac{3m}{\pi^{2}}\left[
\left(  1-\frac{Q}{2}\right)  \int_{0}^{\Lambda}dp\frac{p^{2}}{\sqrt
{\mathbf{p} ^{2}+m^{2}}}+\frac{Q}{2}\int_{0}^{\Lambda}dp\frac{p^{3}
}{\mathbf{p}^{2}+m^{2}}\right]  \label{qq}
\end{equation}
for each quark flavor. At this point we see that the dynamical mass is again
given by a self-consistent equation since the condensate depends also on the
mass. Inserting Eq. (\ref{qq}) into Eq. (\ref{qmdy}) we obtain the deformed
NJL\ gap equation
\begin{equation}
m=\frac{2Gm}{\pi^{2}}\left[  \left(  1-\frac{Q}{2}\right)  \int_{0}^{\Lambda
}dp \frac{p^{2}}{\sqrt{\mathbf{p}^{2}+m^{2}}}+\frac{Q}{2}\int_{0}^{\Lambda}dp
\frac{p^{3}}{\mathbf{p}^{2}+m^{2}}\right]  . \label{qgap}
\end{equation}
It is easy to see that for $Q=0$ $(q=1)$, we recover the NJL gap equation in
its more familiar form
\begin{equation}
m=\frac{2Gm}{\pi^{2}}\int_{0}^{\Lambda}dp \frac{p^{2}}{\sqrt{\mathbf{p}
^{2}+m^{2}}} +m_{0}, \label{gap}
\end{equation}
where $m_{0}$ appears only if we consider the current quark mass term
$\mathcal{L}_{mass}=-m_{0}\overline{\psi}\psi$ in the NJL Lagrangian.

The pion decay constant is calculated from the vacuum to one pion axial vector
current matrix element, which, in the simple 3D non-covariant cutoff we are
using \cite{K}, is given by
\begin{equation}
f_{\pi}^{2}=N_{c} m^{2}\int_{0}^{\Lambda}\frac{d^{3}p}{\left(  2\pi\right)
^{3}} \frac{1}{\left(  \mathbf{p}^{2}+m^{2}\right)  ^{3/2}}, \label{qfpi}
\end{equation}
for each quark flavor. The deformed calculation of $f_{\pi}$ is performed
directly by substituting the dynamical mass in Eq. (\ref{qfpi}) from the one
obtained in Eq. (\ref{qgap}).

As in the non-deformed case, the $q$-gap equation has non-trivial solutions
when the coupling $G$ exceeds a critical value $G_{c}$ related to the cutoff.
The dynamical mass is accordingly modified through the deformed gap equation
(\ref{qgap}).

The behavior of the condensate around the critical coupling, $G_{c},$ is
similar for both deformed and non-deformed cases$,$ meaning that the adopted
procedure to $q$-deform the underlying $su(2)$ algebra in a two flavor NJL
model does not change the behavior of the phase transition around $G_{c}$. The
formalism developed in \cite{ubri,trip} allow $q$-values smaller than one
(which corresponds to $Q>0$). It is worth to mention that in this case the
$q$-deformation effect goes in the opposite direction, namely, the condensate
value and the dynamical mass decrease for $q<1$.

\subsection{Deforming the NJL Hamiltonian}

Here it is important to stress again that the deformed version of this vacuum differs
from the non-deformed one only by a phase and, therefore, the effects of the
deformation comes solely from the deformation of the Hamiltonian.

The deformed functional for the total energy will be obtained from the vacuum
expectation value of the $q$-deformed NJL Hamiltonian:
\begin{equation}
\mathcal{W}^{q}\left[  \theta(p)\right]  =\left\langle NJL\left|
\mathcal{H}_{NJL}^{q}\right|  NJL\right\rangle ,
\end{equation}
where the $q$-deformed Hamiltonian is 
\begin{equation}
\mathcal{H}_{NJL}^{q}=-i\overline{\psi}_{q}\gamma\cdot\nabla\psi_{q}-G\left(
\overline{\psi}_{q}\psi_{q}\right)^2 
-G\left(\overline{\psi}_{q}i\gamma_5\psi_{q}\right)^2\; ,
\end{equation}
and $\psi_{q}$ is given by Eq. (\ref{bd1}). Due to the additive structure of
the $q$-deformation, the deformed Hamiltonian can be written as
\begin{equation}
\mathcal{H}_{NJL}^{q}={H}_{NJL}+\mathcal{H}(Q), \label{HQ}
\end{equation}
and the functional will have the same form
\begin{equation}
\mathcal{W}^{q}\left[  \theta(p)\right]  ={W}\left[  \theta(p)\right]
+\mathcal{W}\left[  Q,\theta(p)\right]  . \label{WQ}
\end{equation}
The last terms of Eqs. (\ref{HQ}) and (\ref{WQ}), namely $\mathcal{H}(Q)$ and 
$\mathcal{W}\left[Q,\theta(p)\right] $, stand for the new terms of first order in $Q$ generated when the algebra is deformed, and, therefore, they vanish for 
$q=1\left(Q=0\right) $. 

The non-deformed case, which corresponds to $q=1\left(  Q=0\right)  $, has been discussed in Sec. II. In order to obtain the new gap equation, we need to calculate the new matrix elements arising from the $q$-deformation of the
NJL Hamiltonian, and add them to the non-deformed functional. This procedure yields to the full $q$-deformed functional for the total energy:
\begin{eqnarray}
{\mathcal W}^{q}\left[ \theta (p)\right]  &=&-2N_{c}N_{f}\left( 1+Q\right)
\int \frac{d^{3}p}{\left( 2\pi \right) ^{3}}p\cos \phi +N_{c}N_{f}\frac{Q}{2}
\int \frac{d^{3}p}{\left( 2\pi \right) ^{3}}p\cos ^{2}\phi   \nonumber \\
&&+N_{c}N_{f}\frac{N_{c}N_{f}}{3\pi ^{2}}QG\Lambda ^{3}\int \frac{d^{3}p}{
\left( 2\pi \right) ^{3}}\cos \phi  
-\left( N_{c}N_{f}\right) ^{2}G\left( 4+Q\right) \left\{ \int \frac{d^{3}p
}{\left( 2\pi \right) ^{3}}\sin \phi \right\} ^{2} \, .
\label{full}
\end{eqnarray}
Defining the new variables
\begin{eqnarray}
P_{q}  & =& P+P_{0}~,\\
P  & =& \left(  1+Q\right)  p~,\\
P_{0}  &=& -\frac{N_{c}N_{f}}{6\pi^{2}}\,G\Lambda^{3}\,Q~,\label{pzero}\\
K &=&-\frac{Q}{2}p~, \\
G^{^{\prime}}  &=& G\left(  1+\frac{Q}{4}\right) \, ,  \label{gprime}
\end{eqnarray}
and performing the same minimization procedure as in the non-deformed case
we obtain
\begin{equation}
P_{q}\ \tan \phi +K\sin \phi =4G^{\prime }N_{c}N_{f}\int 
\frac{d^{3}p^{\prime }}{\left( 2\pi \right) ^{3}}\sin \phi \, .  
\label{qbvgap}
\end{equation}
The term $K\sin\phi$ was overlooked in a previous paper \cite{CLL} \footnote{Figure 1 of Ref. \cite{CLL} shows the variational angle as a function of $q$ when the dynamical mass is equal to $p$, not 300 MeV as stated in the caption.}. 
Trying to keep as much similarity as possible with the usual solution, instead of
solving the equation $P_{q}\tan\phi (p)+K\sin\phi (p)  =M$,
the $\sin\phi (p) $ will be substituted by $\alpha\tan\phi (p)$, with $\alpha = 0.59$ chosen to minimize the difference between
$\sin\phi$ and $\tan\phi$ in the interval $[0,\pi/4]$.

The new gap equation then becomes
\begin{equation}
M=4G^{^{\prime}}N_{c}N_{f}\int\frac{d^{3}p}{\left(  2\pi\right)  ^{3}}
\frac{M}{\sqrt{\mathbf{K}_{q}^{2}+M^{2}}}\,, 
\label{qgap1}
\end{equation}
provided the variational angles have the same old structure (but they now are
$q$-dependent)
\begin{equation}
\tan2\theta_{q}(p)=\frac{M}{K_{q}},\sin2\theta_{q}(p)=\frac{M}{\sqrt
{\mathbf{K}_{q}^{2}+M^{2}}},\cos2\theta_{q}(p)=\frac{K_{q}}{\sqrt
{\mathbf{K}_{q}^{2}+M^{2}}} \, ,
\label{qang}
\end{equation}
where
\begin{equation}
K_q = P_q + \alpha K \, .
\end{equation}

It is easy to see that, when $q\rightarrow1\left(  Q\rightarrow0\right)  $,
Eqs. (\ref{qbvgap}), (\ref{qgap1}), and (\ref{qang}) reduce to their
non-deformed versions, since
$K_q\rightarrow p$ and $G^{^{\prime}}\rightarrow G$.

In analogy with the non-deformed case we can write the gap equation in terms
of the quark condensates as
\begin{equation}
M=-2G^{^{\prime}}\left\langle \overline{\Psi}\Psi\right\rangle .
\label{gapqq}
\end{equation}
Comparing the two forms of the gap equation Eqs. (\ref{qgap}) and
(\ref{gapqq}), we find a new deformed condensate given by
\begin{equation}
\left\langle \overline{\Psi}\Psi\right\rangle =-\frac{N_{c}N_{f}}{\pi^{2}}\int
_{0}^{\Lambda}dp\,p^{2}\frac{M}{\sqrt{\mathbf{K}_{q}^{2}+M^{2}}}
\label{qqdef}
\end{equation}
This condensate is different from the one obtained in the previous section,
where the condensate was explicitly deformed \cite{VST}. 
It also has exactly the same form of the non-deformed one, but is
written in terms of the new variables. It is worth to mention that the new
condensate is not obtained by calculating the vacuum expectation value of a
deformed scalar density. In fact, it corresponds to the gap equation
which arises from the variational procedure started from the $q$-deformed
Hamiltonian. The new pion decay constant can also be obtained in analogy to
the non-deformed case
\begin{equation}
F_{\pi}^{2}=N_{c}M^{2}\int_{0}^{\Lambda}\frac{d^{3}p}{\left(  2\pi\right)
^{3}}\frac{1}{\left(  \mathbf{K}_q^2 + M^2\right)^{3/2}}.
\label{FPI}
\end{equation}

A closer look to the left side of Eq. \ref{qbvgap} shows that the dynamical
mass has two components: one proportional to the original dynamical mass
$m$ and another term which we call $M_{0}$
\begin{equation}
M=\underbrace{\left[1+(1-\alpha /2)Q\right]
\underbrace{p\,\tan2\theta(p)}_{m}}_{M_{dyn}}+
\underbrace{p_{0}\tan2\theta(p)}_{M_{0}}\,, 
\label{qgapqa}
\end{equation}
where $p_{0}=-\frac{QG\Lambda^{3}}{\pi^{2}}$ is $P_{0}$ (Eq. \ref{pzero}) for
$N_{c}=3$ and $N_{f}=2$. 
We can then calculate the effect of the deformation on the dynamical mass $m$: 
\begin{equation}
m=\frac{M}{1+(1-\alpha /2)Q}\, ,
\label{mf}
\end{equation}
where $M$ is obtained by solving Eq. (\ref{gapqq}). In an analogous way, the
effect on the condensate can be calculated by substituting $M\rightarrow
m$ in the numerator of Eq. (\ref{qqdef})
\footnote{The term $M_0$ has to be considered only when solving the gap equation. But it can be neglected when going from Eq. (\ref{qgapqa}) to (\ref{mf}), since it is much smaller than the dynamical mass, $M$.} .

\section{Temperature and $q$-Deformation}

In this section we want to study the interplay of temperature and
$q$-deformation in the NJL model. We review the standard approach to introduce finite temperature and
chemical potential in the NJL model \cite{ASYA,VW} in subsections IV.A and IV.B. When temperature is introduced in the model, the condensate is replaced by the thermal
expectation value of the scalar density, which contains the Fermi-Dirac distributions. We also have a gap equation for the chemical potential, which is modified by the interactions in such a way that we need to solve a system of coupled gap equations. In subsections IV.C and IV.D, the standard formalism is extended to incorporate the effects of $q$-deformation. In particular, aspects of chiral symmetry restoration and pionic properties in the $q<1$ regime will be discussed.

\subsection{The NJL model at Finite Temperature}

The starting point for the study of the thermodynamics of the NJL model is the
partition function
\begin{equation}
\mathcal{Z}=Tr\exp\left[  -\beta\left(  \mathcal{H}_{NJL}-\mu_{i}
\mathcal{N}_{i}\right)  \right]  ,
\label{pfz}
\end{equation}
where $\beta=T^{-1}$, $\mathcal{N}_{i}$ is the valence quark number operator
of flavor $i$, and $\mu_{i}$ are the corresponding chemical potential. From
the above partition function, one can calculate the thermal expectation value
of an operator $\mathcal{A}$
\begin{equation}
\left\langle \left\langle \mathcal{A}\right\rangle \right\rangle
=\frac{1}{\mathcal{Z}}Tr\mathcal{A}\left\{  \exp\left[  -\beta\left(
\mathcal{H}_{NJL}-\mu_{i}\mathcal{N}_{i}\right)  \right]  \right\},
\label{txv}
\end{equation}
where the operator in question can be, for example, the scalar density
$\overline{\psi}\psi$ or the quarks density $\psi^{\dagger}\psi$.

Here the Hamiltonian is given by
\begin{align}
\mathcal{H}_{NJL} &  =-i\overline{\psi}\gamma\cdot\nabla\psi-G\left(
\overline{\psi}\psi\right)  ^{2}-G\left(  \overline{\psi}i\gamma
_{5}\mathbf{\tau}\psi\right)  ^{2}+m_{0}\overline{\psi}\psi,\label{HMASS}\\
&  =-i\overline{\psi}\gamma\cdot\nabla\psi-\mathcal{L}_{int}+m_{0}
\overline{\psi}\psi,
\end{align}
which, in the mean field approximation, is written as
\begin{equation}
\mathcal{H}_{MF}=-i\overline{\psi}\gamma\cdot\nabla\psi+m\overline{\psi}
\psi+G\sigma_{1}^{2}+\frac{G}{N_{c}}\sigma_{2}\mathcal{N}-\frac{G}{2N_{c}
}\sigma_{2}^{2},\label{HMFT}
\end{equation}
where $m$ is the effective quark mass
\begin{equation}
m=m_{0}-2G\sigma_{1},\label{gapm}
\end{equation}
$\mathcal{N}$ is the number operator
\begin{equation}
\mathcal{N}=\overline{\psi}\gamma_{0}\psi=\psi^{\dagger}\psi,
\end{equation}
and $\sigma_{1}$, $\sigma_{2}$ are defined as
\begin{align}
\sigma_{1} &  =\left\langle \left\langle \overline{\psi}\psi\right\rangle
\right\rangle ,\\
\sigma_{2} &  =\left\langle \left\langle \overline{\psi}\gamma_{0}
\psi\right\rangle \right\rangle .
\end{align}
The Hamiltonian in the mean field approximation represents a system of free
particles of mass $m$ and the chemical potential is given by
\begin{equation}
\mu=\mu_{0}-\frac{G}{N_{c}}\sigma_{2},\label{gapmu}
\end{equation}
where $\mu_{0}$ is the chemical potential when there is no vector interaction,
generated by the Fierz transformation of $\mathcal{L}_{int}$. Equations
(\ref{gapm}) and (\ref{gapmu}) form a system of self-consistent coupled
equations
\begin{equation}
\left\{
\begin{array}
[c]{l}
m=m_{0}-2G\sigma_{1},\\
\mu=\mu_{0}-\frac{G}{N_{c}}\sigma_{2}.
\end{array}
\right.
\end{equation}
Solving the above equations we obtain the effective mass and chemical
potential, which we can put back in the expressions for $\sigma_{1}$ and
$\sigma_{2}$ to obtain the condensate and the density at a given temperature
$T$.

\subsection{Thermal Expectation Values}

The dynamical mass calculation depends on expectation values containing the
Fermi-Dirac distributions. The thermal expectation values relevant for the NJL
model, namely $\sigma_{1}$ and $\sigma_{2}$, can be calculated within the
formalism of the thermal Green function \cite{dolan}, which has the advantage
of separating the parts that depend on the temperature and the chemical
potential. The thermal Green function for a free fermion at a temperature $T$
and chemical potential $\mu$ is written as
\begin{align}
S_{F}\left(  p;T,\mu\right)   &  =\frac{p\kern  -.5em\slash  +m}{p^{2}
-m^{2}+i\varepsilon}+2\pi i\delta\left(  p^{2}-m^{2}\right) \nonumber\\
&  \times\left(  p\kern  -.5em\slash  +m\right)  \left[  \theta\left(
p^{o}\right)  n(\mathbf{p},T,\mu)+\theta\left(  -p^{o}\right)  \overline
{n}(\mathbf{p},T,\mu)\right]  , \label{SFT}
\end{align}
where
\begin{equation}
n(\mathbf{p},T,\mu)=\frac{1}{1+\exp\left[  -\beta\left(  E-\mu\right)
\right]  }, \label{nf}
\end{equation}
and
\begin{equation}
\overline{n}(\mathbf{p},T,\mu)=\frac{1}{1+\exp\left[  -\beta\left(
E+\mu\right)  \right]  }, \label{naf}
\end{equation}
are the fermions and anti-fermions distribution function respectively with
$E=\sqrt{\mathbf{p}^{2}+m^{2}}$.

Making use of the anti-commutation relation for the fermionic fields, the
expectation values can be written as
\begin{align}
\left\langle \left\langle \overline{\psi}(x)\psi(x)\right\rangle
\right\rangle  &  =-Tr\lim_{y\rightarrow x+}\left\langle \left\langle T\left[
\psi(x)\overline{\psi}(y)\right]  \right\rangle \right\rangle ,\\
\left\langle \left\langle \overline{\psi}\gamma_{0}\psi\right\rangle
\right\rangle  &  =-Tr\lim_{y\rightarrow x+}\gamma_{0}\left\langle
\left\langle T\left[  \psi(x)\overline{\psi}(y)\right]  \right\rangle
\right\rangle .
\end{align}
The term on the right-hand side of the above equations is the definition of
the thermal Green function in the configuration space:
\begin{equation}
iS_{F}\left(  x-y;T,\mu\right)  =\left\langle \left\langle T\left[
\psi(x)\overline{\psi}(y)\right]  \right\rangle \right\rangle .
\end{equation}
The expectation values are then re-written as
\begin{align}
\left\langle \left\langle \overline{\psi}(x)\psi(x)\right\rangle
\right\rangle  &  =-i\lim_{y\rightarrow x+}Tr\int\frac{d^{4}p}{\left(
2\pi\right)  ^{4}}S_{F}\left(  p;T,\mu\right)  e^{-i(x-y)p},\\
\left\langle \left\langle \overline{\psi}\gamma_{0}\psi\right\rangle
\right\rangle  &  =-i\lim_{y\rightarrow x+}Tr\int\frac{d^{4}p}{\left(
2\pi\right)  ^{4}}\gamma_{0}S_{F}\left(  p;T,\mu\right)  e^{-i(x-y)p}.
\end{align}
Applying a cutoff in the momentum, the results for $\sigma_{1}$ and
$\sigma_{2}$ are the following
\begin{align}
\sigma_{1} &  =\left\langle \left\langle \overline{\psi}\psi\right\rangle
\right\rangle =-\frac{N_{c}N_{f}}{\pi^{2}}\int_{0}^{\Lambda}dpp^{2}\frac{m}
{E}\left[  1-n-\overline{n}\right]  ,\\
\sigma_{2} &  =\left\langle \left\langle \overline{\psi}\gamma_{0}
\psi\right\rangle \right\rangle =\frac{N_{c}N_{f}}{\pi^{2}}\int_{0}^{\Lambda
}dpp^{2}\left[  n-\overline{n}\right]  ,
\end{align}
and the barion density can be written as
\begin{equation}
\rho_{B}=\frac{1}{3}\left\langle \left\langle \overline{\psi}\gamma_{0}
\psi\right\rangle \right\rangle =\frac{N_{c}N_{f}}{3\pi^{2}}\int_{0}^{\Lambda
}dpp^{2}\left[  n-\overline{n}\right]  .
\end{equation}

\subsection{Condensates, mass, and chiral symmetry restoration}

An interesting window to the application of $q$-deformation to hadronic physics can
be opened for $q$-values smaller than one. It is tempting to explore the
behavior of the condensate in this new regime for smaller values of
$q$, even considering that the truncation at order $Q$ may not be granted. In
this case, we observed in \cite{CLL} that the chiral symmetry is restored in
the limit $q\rightarrow0$, since the condensate vanishes. It is
then worth to investigate the effect of both temperature and $q$-deformation
in the chiral symmetry restoration process as well as in the pion properties.

In order to study the effect of both temperature and $q$-deformation in the
chiral symmetry restoration process, we can replace the gap equation for the mass in the system of coupled equations by its deformed version Eq. (\ref{qgapm})
\begin{equation}
M=m_{0}-2G^{\prime}\left\langle \left\langle \overline{\Psi}\Psi\right\rangle
\right\rangle \;,
\label{qgapm}
\end{equation}
so that the new system of coupled equation we need to solve is
\begin{equation}
\left\{
\begin{array}
[c]{l}
M=m_{0}-2G^{\prime}\Sigma_1,\\
\mu=\mu_{0}-\frac{G}{N_{c}}\sigma_{2},
\end{array}
\right.
\end{equation}
where $\Sigma_1$ represents the deformed version of the thermal condensate $\sigma_1$.

Solving the new set of coupled gap equations we can observe the condensate as
a function of temperature for different values of the deformation parameter.
It is important to mention that the effects of the algebra deformation come
from gap equation for the mass. However, the chemical potential is also
affected once it depends on the mass through the coupled equations. 
Thus, for a given temperature, we obtain the mass and the chemical potential 
by solving the system of coupled gap equations. We then calculate the condensate 
and the density.

\begin{figure}[b]
\centering
\includegraphics[width=7cm ,height=7cm]{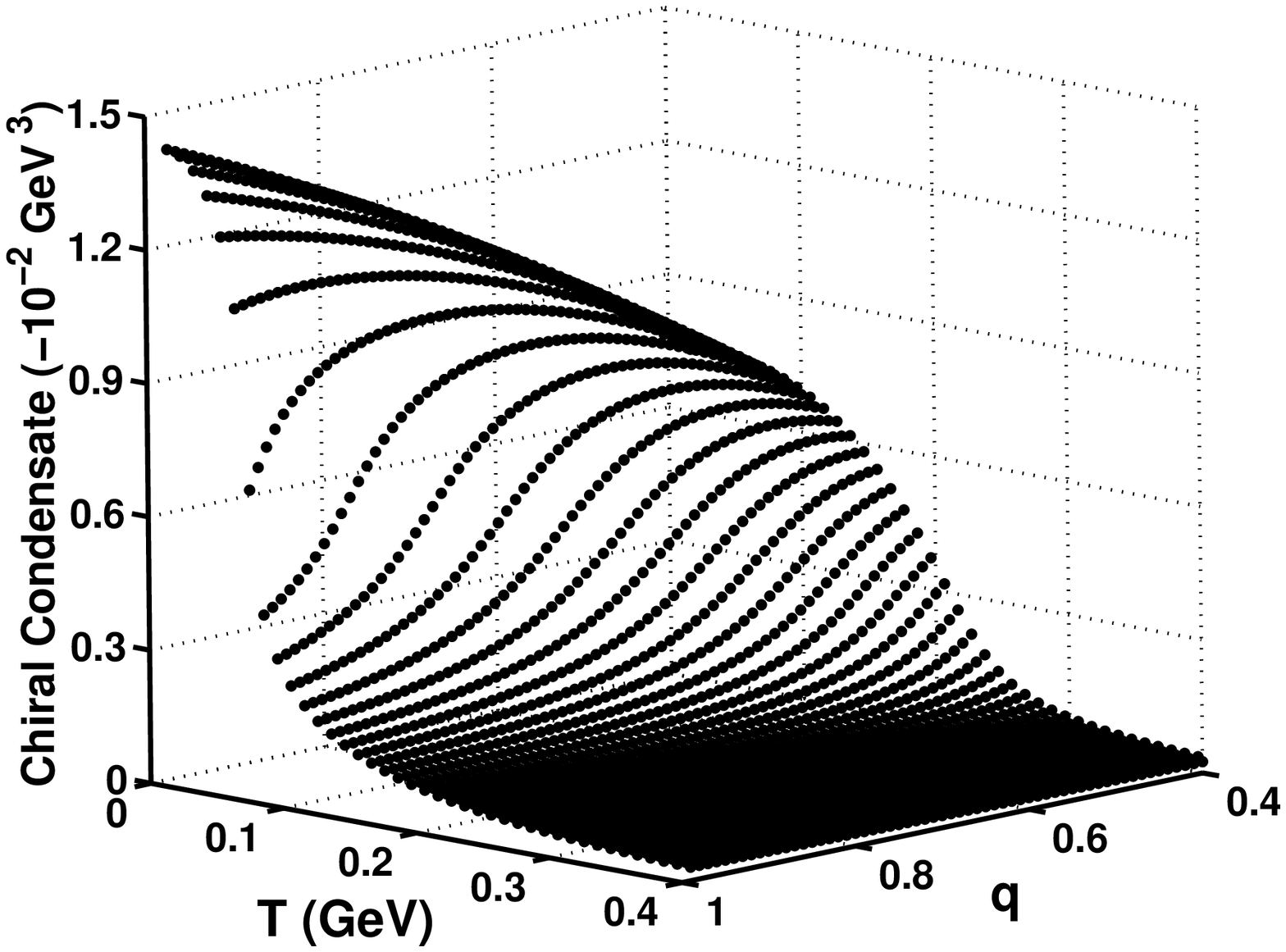}
\hspace{0.5cm}
\includegraphics[width=7cm ,height=7cm]{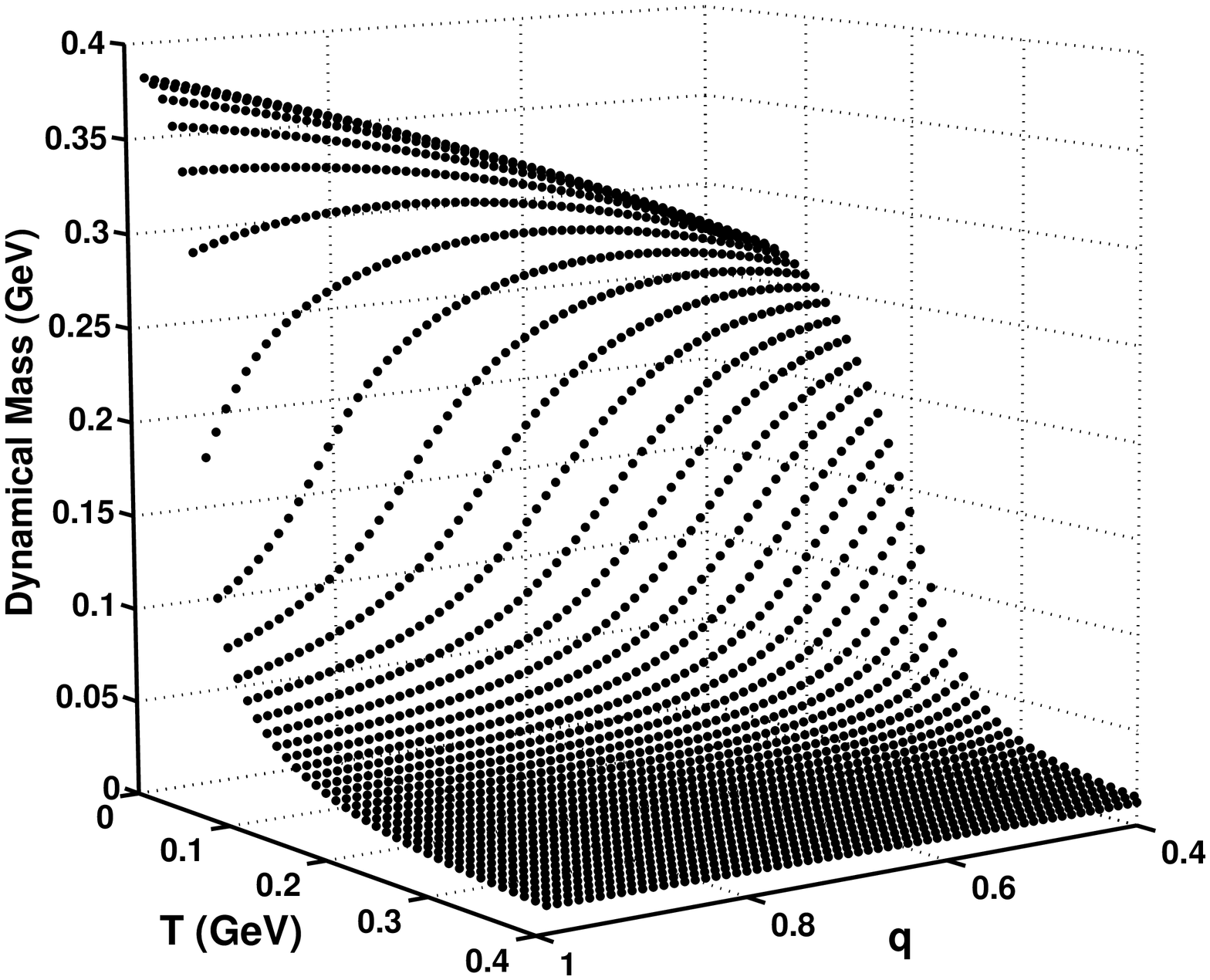}
\caption{The chiral condensate and the dynamical quark mass as functions of both temperature and $q$-deformation in the $q<1$ regime. The results presented in the
figures were obtained with the following parameters: 
$G=6.58$ GeV$^{-2}$, 
$\Lambda = 0.6$ GeV, 
$m_0=5.87$ MeV, and 
$\mu_0=0.35$ GeV.}
\label{CCmass}
\end{figure}

In order to visualize the chiral symmetry restoration process with the
influence from both temperature and deformation, we plot the surfaces shown in
Fig. \ref{CCmass} where the chiral condensate and the constituent quark mass 
are plotted as functions of both temperature and $q$-deformation. 

At $q < 1$ deformations, an interesting result is observed: the chiral symmetry restoration is similar to the one observed by Klimt {\it et al.} \cite{klimt} as function of temperature, $T$, and density, $\rho$. However, the process is slower as $q$ gets smaller when compared with the case of $\rho$ getting larger. Figure \ref{CCmass} shows that the condensate is substantially affected by the deformation and lowers as $q$ decreases, while the effect of temperature is softened. The same behavior is observed in the dynamical quark mass since it is linearly constrained to the condensate by the gap equation. 

\begin{figure}[t]
\centering
\includegraphics[width=7cm ,height=7cm]{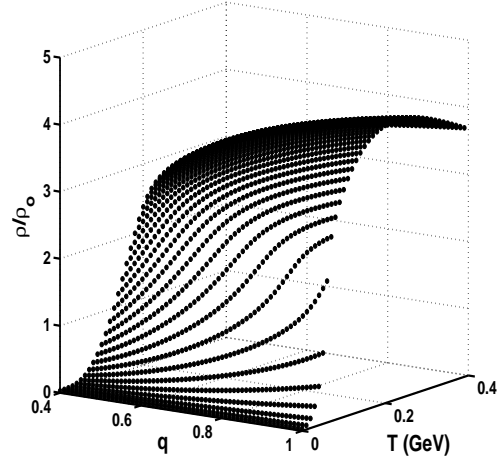}
\caption{The barion density in units of the nuclear matter density $\rho_0 = 0.15 
fm^{-3}$ as a function of both temperature and $q$-deformation in the $q<1$ regime.}
\label{Rho}
\end{figure}

We have also calculated the barion density as a function of $T$ and $q$, which is displayed in Fig. \ref{Rho}. The effect of the deformation on the barion density in only observed as the temperature increases. The density also lowers as $q$ decreases, and the temperature effect is also smoothed down. 

The behavior of the dynamical mass and the chiral condensate directs to chiral symmetry restoration. We would therefore expect the pion to become massless in the limit $q \rightarrow 0$. In the next section, we evaluate the pion decay constant and mass in order to investigate the pion properties in a regime where condensate is lower than in the standard NJL model due to the effects of the $q$-deformation.

\subsection{Pion Properties in the $q<1$ regime}

Recent works \cite{stern} have shown that the spontaneous chiral symmetry breaking
may result from a balance of the density of states and their
mobility in the medium. It may occur with a vanishing $<\bar{q}q>$ condensate
as long as the low density of states is compensated by a high mobility. In
this case, the condensate is no longer the order parameter.

The effective models for QCD, like the NJL, fail to describe hadronic
observables when the condensate is small or null. The $q$-deformed version of
the NJL model seems to be an alternative to study the low condensation
scenario. As a first application, we start with the pion mass since it cannot be reproduced in the standard NJL model in a low condensation regime.

\begin{figure}[b]
\centering
\includegraphics[width=7cm ,height=7cm]{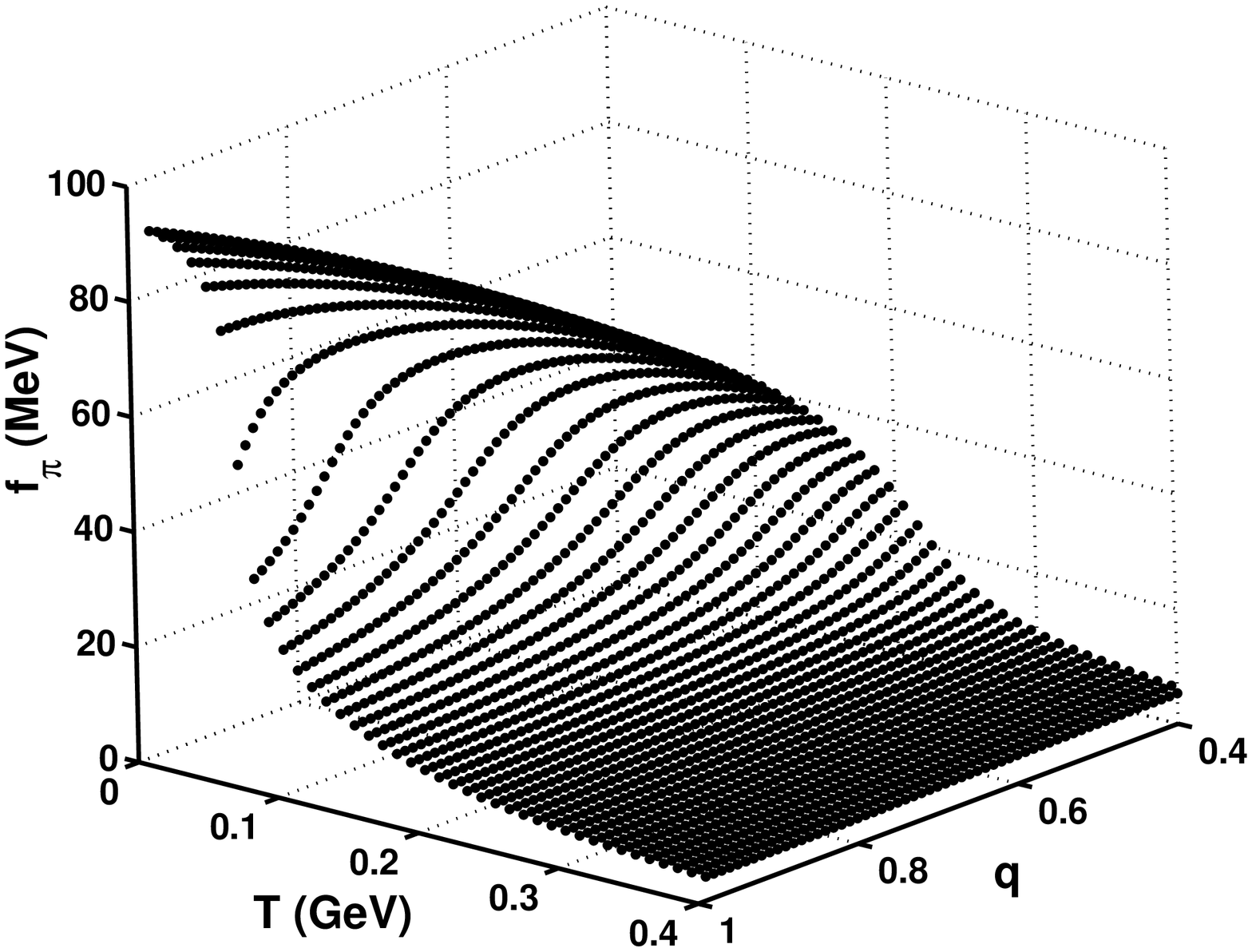}
\hspace{0.5cm}
\includegraphics[width=7cm ,height=7cm]{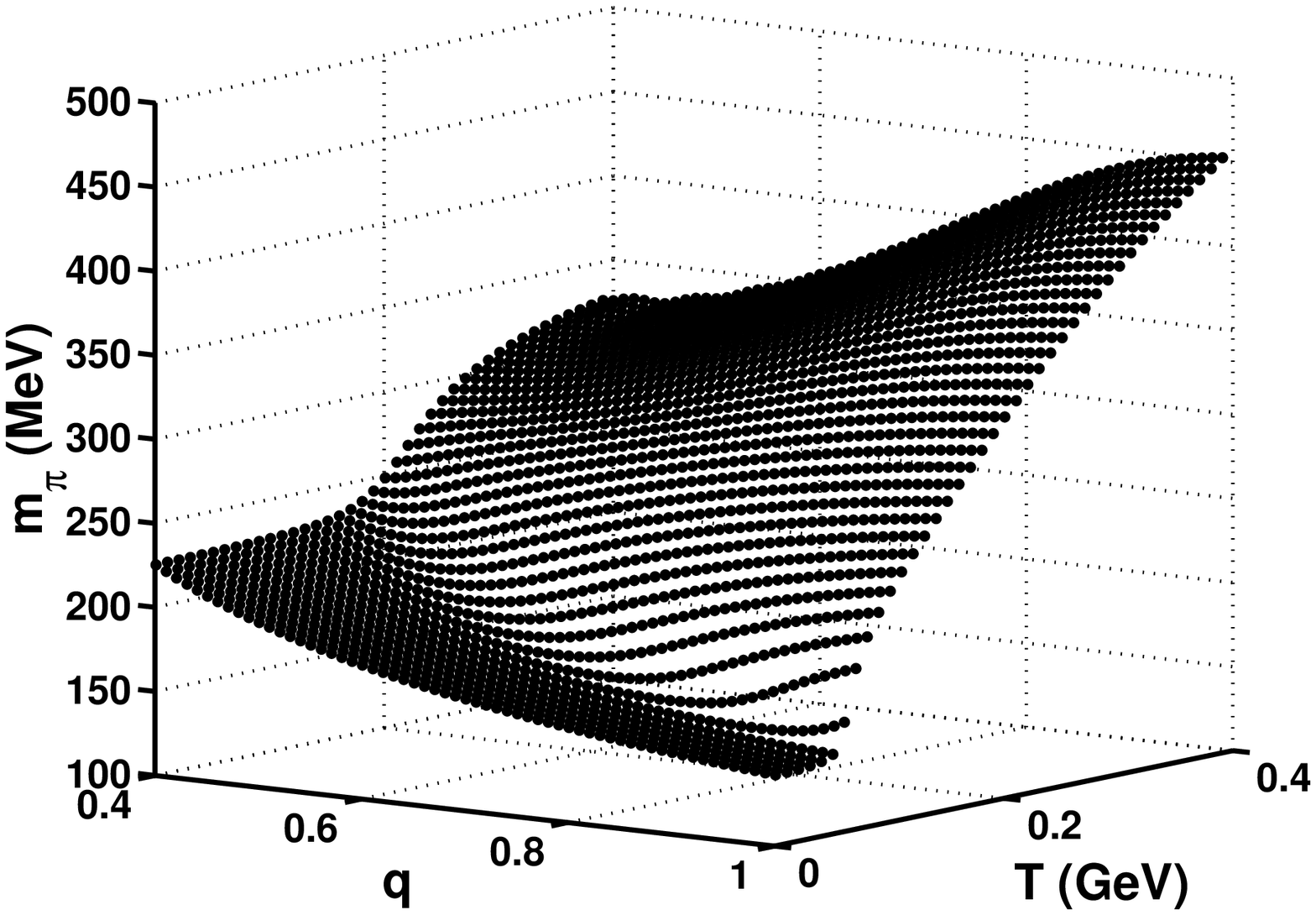}
\caption{The pion decay constant and the pion mass as functions of both temperature and 
$q$-deformation in the $q<1$ regime.}
\label{FpiMpi}
\end{figure}

With a fixed value for the current quark mass, after obtaining the pion decay constant through Eq. \ref{FPI}, we obtain the pion mass by using the Gell-Mann-Oakes-Renner (GOR) relation 
\begin{eqnarray}
m_{\pi}^{2}{F_{\pi}^{2}}=\left(m^u_0+m^d_0\right)\times{\left|  \left\langle \bar{\Psi
}\Psi\right\rangle \right|  }\; ,
\label{GOR}
\end{eqnarray}
written in terms of our deformed quantities. The pion properties (decay constant and mass) as functions of both temperature and $q$-deformation are shown in Figure \ref{FpiMpi}. 
In order to have an idea of the effect of the deformation in a low condensation regime, we consider the case of $q=0.5$, where condensation is approximately 35\% lower than in the non-deformed case (see Fig. \ref{CCmass}). In this particular scenario, we obtain a pion mass of about 190 MeV. The standard NJL model would give a pion mass of about 100 MeV in such condensation scenario. The correlations between the constituents of the system introduced when the underlying $su(2)$ algebra is deformed seems to compensate the low condensation, which is responsible for keeping a reasonable mass for the pion even when the value of the condensate is getting smaller. Recent measurements of the semileptonic decay $K^+\rightarrow\pi^+\pi^-e^+\nu_e$ \cite{pislak} obtained a value of the $\pi\pi$ $S$-wave scattering length which indicates that the condensate is indeed the order parameter \cite{colangelo}. However, it is interesting to note that the deformed version of the NJL model is able to describe the pion  even with smaller values of the condensate.

\section{Final Remarks}

So far we have performed the $q$-deformation of the NJL model with finite
temperature also taken into account. We studied the $q$-deformation of the underlying $su(2)$ algebra in a two flavor version of the NJL model and investigated an important feature of chiral symmetry: its restoration at finite temperature. Also, we studied the pion properties in a low condensation scenario, where the standard NJL model would underestimate the pion mass. 

As far as the effects of both temperature and deformation are concerned, our main conclusions can be summarized as follows. The quark condensate and the dynamical mass decrease as $q$ gets smaller than one, and chiral symmetry would be restored in the limit $q\rightarrow 0$ \cite{CLL}. This effect is specially observed at low temperatures, since we have thermal restoration at large $T$.  Surprisingly, as the condensate lowers due to the deformation, the pion not only keeps massive but also increases. The GOR relation, Eq. \ref{GOR}, with our deformed quantities gives the reason for this behavior, where the pion decay constant decreases faster than the square root of the condensate.

It seems that the effect of  the deformation compensates the lower condensation, suggesting that the quarks' mobility is affected when the algebra is deformed. In fact, the correlations introduced by the deformation generate an extra momentum $P_0$ for the constituents of the system (see Eq. \ref{pzero}) which is responsible for enhancing the mobility in the medium.
\newline \newline
\textbf{Acknowledgments}
\newline\newline
The authors are grateful to U-G. Mei\ss ner and M. R. Robilotta for the
suggestion which motivated our study on the $m_\pi$ and $f_{\pi}$ behavior, 
and to D. Galetti and B. M. Pimentel for very helpful discussions. This work was
supported by FAPESP grant number 2002/10896-7. V.S.T would like to thank
FAEPEX/UNICAMP for partial financial support.




\end{document}